\documentclass[aps,prd,superscriptaddress,showpacs,preprint]{revtex4}
\usepackage{graphicx, bm}
\usepackage[usenames]{color}
\usepackage{float}
\usepackage{multirow}
\usepackage{subcaption}

\begin{document}

\draft
\title{Probing the heavy Higgs boson production and decay $H_0$ of the  Bestest Little Higgs Model at the LHC and the FCC-hh}

\author{E. Cruz-Albaro\footnote{elicruzalbaro88@gmail.com}}
\affiliation{\small Facultad de F\'{\i}sica, Universidad Aut\'onoma de Zacatecas\\
            Apartado Postal C-580, 98060 Zacatecas, M\'exico.\\}

\author{A. Guti\'errez-Rodr\'{\i}guez\footnote{alexgu@fisica.uaz.edu.mx}}
\affiliation{\small Facultad de F\'{\i}sica, Universidad Aut\'onoma de Zacatecas\\
      Apartado Postal C-580, 98060 Zacatecas, M\'exico.\\}



\author{D. Espinosa-G\'omez \footnote{david.espinosa@umich.mx}
}
\affiliation{\small Facultad de Ciencias F\'{\i}sico Matem\'aticas, Universidad Michoacana de San Nicol\'as de Hidalgo\\
            Avenida Francisco, J. M\'ujica S/N, 58060, Morelia, Michoac\'an, M\'exico.\\}

\author{T. Cisneros-P\'erez \footnote{tzihue@gmail.com}
}
\affiliation{\small Unidad Acad\'emica de Ciencias Qu\'{\i}micas, Universidad Aut\'onoma de Zacatecas\\
         Apartado Postal C-585, 98060 Zacatecas, M\'exico.\\}

\author{F. Ramírez-Zavaleta \footnote{feryuphy@yahoo.com.mx}
}
\affiliation{\small Facultad de Ciencias F\'{\i}sico Matem\'aticas, Universidad Michoacana de San Nicol\'as de Hidalgo\\
            Avenida Francisco, J. M\'ujica S/N, 58060, Morelia, Michoac\'an, M\'exico.\\}


\date{\today}

\begin{abstract}

In the Bestest Little Higgs Model (BLHM) scenario, we analyze the branching ratios and production cross-section of the heavy Higgs boson $H_0$. The analysis is performed at the tree level and the one-loop level. In addition, we present results of the possible production of the heavy Higgs boson $H_0$ via gluon fusion for the center-of-mass energies and the integrated luminosities of the LHC, HE-LHC, HL-LHC, and FCC-hh. Our results show a very optimistic scenario for studying the $H_0$ scalar predicted by the BLHM and for the energies and luminosities of current and future hadron colliders.

\end{abstract}


\vspace{5mm}

\maketitle

\section{Introduction}

There are convincing theoretical arguments and a wide range of experimental facts that motivate the need for new physics beyond the Standard Model (SM), such as the hierarchy problem, the strong CP problem, the baryon asymmetry of the universe, the existence of dark matter, the fine-tuning of the Higgs boson mass, the origin of fermionic families, etc.. Most of the solutions to these problems require new interactions and new particles, such as supersymmetric partners, heavy Higgs bosons, dark photons, axions, right-handed neutrinos, and monopoles, among other things.

Many of the proposed new physics models contain an extended Higgs sector, among which the BLHM~\cite{JHEP09-2010,Godfrey:2012tf,Kalyniak,Cisneros-Perez:2023foe,Cruz-Albaro:2023pah,Cruz-Albaro:2022lks,Cruz-Albaro:2022kty} is one of the viable options because it provides an exciting way to address the hierarchy problem without resorting to fine-tuning. In addition, it solves some issues that are present in the great majority of the other Little Higgs models (The Littlest Higgs Model~\cite{Arkani-Hamed:2002ikv}, A Littlest Higgs Model with custodial $SU(2)$ symmetry~\cite{Chang:2003zn}, The Little Higgs Model~\cite{Han:2003wu}, The Little Higgs Model and custodial $SU(2)$~\cite{Chang:2003un}, The Simplest Little Higgs Model~\cite{Schmaltz:2004de}), such as the problem of dangerous singlets~\cite{Schmaltz:2008vd}, a pathology where collective symmetry breaking does not suppress quadratically divergent corrections to the Higgs boson mass; and strong constraints from precision electroweak observables~\cite{Han:2013ic,Reuter:2012sd} in the gauge sector.
Instead, the BLHM generates a viable Higgs quartic coupling where the real singlet $\sigma$ is no longer a dangerous singlet; that is, it no longer develops a divergent tadpole from radiative corrections~\cite{JHEP09-2010}. On the other hand, a custodial SU(2) symmetry~\cite{Csaki:2002qg} and disassociation in the masses of the new quarks and heavy gauge bosons are implemented in the BLHM, thus avoiding the constraints from precision measurements.

The disassociation in the masses of the partners of fermions ($ T $, $ B $, $ T_5 $, $ T_6 $, $ T^{2/3} $, $ T^{5/3} $) and gauge bosons ($ Z' $, $ W'^{\pm} $) is achieved by incorporating two independent symmetry-breaking scales, $ f $ and $ F $ with $F>f$. This leads to new quarks with masses proportional only to the $ f $ scale, while the new gauge bosons acquire masses proportional to the combination of the $ f $ and $ F $ scales. Since the new quarks are now lighter than the new gauge bosons, fine-tuning in the top sector and electroweak precision constraints in the gauge sector are avoided.
In the scalar sector, neutral and charged physical scalar fields also arise: $h_0, H_0, A_0, \phi^{0},\eta^{0}, H^{\pm}, \phi ^{\pm}$ and $ \eta^{\pm}$.  The $h_0$ state is assumed to be light ($ \approx 125 $ GeV), similar to the Higgs of the SM, while the rest of the scalars are allowed to vary their masses.

In this paper, we explore the production of the heavy Higgs boson $H_0$  of the BLHM at current and future colliders such as the Large Hadron Collider (LHC) and the Future Circular Collider hadron-hadron (FCC-hh)~\cite{Abada:2019FCC2}, respectively.
The discovery of any Higgs boson beyond the SM will be unequivocal evidence for the existence of an extended Higgs sector.  Therefore, probing Higgs sectors of extended models through direct searches for new Higgs bosons at high-energy colliders or through modifications to SM-like Higgs couplings tested by precision measurements of Higgs coupling take on a transcendental role~\cite{ATLAS:2019slw}.  The reason for this is that they open new routes to explore new physics effects.

Until now, direct and indirect searches for new physics at a weak scale have produced only unsuccessful results. However, new physics is expected to emerge at high masses, which means that it will be necessary to maximize the center-of-mass energy $ \sqrt{s} $ of current colliders so that new heavy particles can be produced in collisions. The mentioned above motivates the construction of more energetic colliders with higher luminosity $ \mathcal{L} $. Therefore, the search for new physics beyond the SM remains a frontier in particle physics research.

The article is organized as follows. In Section~\ref{BLHM}, we briefly review the BLHM. In Section~\ref{H0decay}, we present and study the tree-level and one-loop decays of the  Higgs boson $ H_0 $. In Section~\ref{results}, we show the predictions of the BLHM in the production cross-sections of the $ H_0 $ Higgs for the processes via gluon fusion. Finally, conclusions are presented in Section~\ref{conclusions}.

\section{A brief review of the BLHM} \label{BLHM}

The BLHM~\cite{JHEP09-2010,Godfrey:2012tf,Kalyniak,Cisneros-Perez:2023foe,Cruz-Albaro:2023pah,Cruz-Albaro:2022lks,Cruz-Albaro:2022kty} is based on two independent non-linear sigma models. With the first field $\Sigma$, the global symmetry
$SO(6)_A\times SO(6)_B$ is broken to the diagonal group $SO(6)_V$ at the energy scale $f$, while with the second field
$\Delta$, the global symmetry $SU(2)_C \times SU(2)_D$ to the diagonal subgroup $SU(2)$ to the scale $F> f$. In the first stage, 15 pseudo-Nambu-Goldstone bosons that are parameterized as

\begin{equation}\label{Sigma}
\Sigma=e^{i\Pi/f}  e^{2i\Pi_{h}/f}e^{i\Pi/f},
\end{equation}

\noindent
where $\Pi$ and $\Pi_h$ are complex and antisymmetric matrices given in Ref.~\cite{JHEP09-2010}. Regarding the second stage of spontaneous symmetry-breaking, the pseudo-Nambu-Goldstone bosons of the field $\Delta$ are parameterized as follows

\begin{equation}\label{Delta}
\Delta=F e^{2i \Pi_d/F},\, \,\, \, \, \Pi_d=\chi_a \frac{\tau^{a}}{2} \ \ (a=1,2,3),
\end{equation}

\noindent
$\chi_a$ represents the Nambu-Goldstone fields and the $\tau_a$ correspond to the Pauli matrices~\cite{JHEP09-2010}, which are the
generators of the SU(2) group.

\subsection{The scalar sector}

The BLHM Higgs fields, $h_1$ and $h_2$, form the Higgs potential that undergoes spontaneous symmetry breaking~\cite{JHEP09-2010,Kalyniak,Erikson}:

\begin{equation}\label{Vhiggs}
V_{Higgs}=\frac{1}{2}m_{1}^{2}h^{T}_{1}h_1 + \frac{1}{2}m_{2}^{2}h^{T}_{2}h_2 -B_\mu h^{T}_{1} h_2 + \frac{\lambda_{0}}{2} (h^{T}_{1}h_2)^{2}.
\end{equation}

\noindent  The symmetry-breaking mechanism is implemented in the BLHM when the Higgs doublets acquire
their vacuum expectation values (VEVs), $\langle h_1\rangle ^{T}=(v_1,0,0,0)$ and $ \langle h_2 \rangle ^{T}=(v_2,0,0,0)$. By demanding that
these VEVs minimize the Higgs potential of Eq.~(\ref{Vhiggs}), the following relations are obtained

\begin{eqnarray}\label{v12}
&&v^{2}_1=\frac{1}{\lambda_0}\frac{m_2}{m_1}(B_\mu-m_1 m_2),\\
&&v^{2}_2=\frac{1}{\lambda_0}\frac{m_1}{m_2}(B_\mu-m_1 m_2).
\end{eqnarray}

\noindent These parameters can be expressed as follows

\begin{equation}\label{vvacio}
v^{2}\equiv v^{2}_1 +v^{2}_2= \frac{1}{\lambda_0}\left( \frac{m^{2}_1 + m^{2}_2}{m_1 m_2} \right) \left(B_\mu - m_1 m_2\right)\simeq \left(246\ \ \text{GeV}\right)^{2},
\end{equation}

\begin{equation}\label{beta}
\text{tan}\, \beta=\frac{v_1}{v_2}=\frac{m_2}{m_1}.
\end{equation}

\noindent From the diagonalization of the mass matrix for the scalar sector,
three non-physical fields $G_0$ and $G^{\pm}$, two physical scalar fields $H^{\pm}$
and three neutral physical scalar fields $h_0$, $H_0$ and $A_0$ are generated~\cite{Kalyniak,PhenomenologyBLH}. The lightest state, $h_0$, is identified as the scalar boson of the SM.
The four parameters in the Higgs potential $ m_1,  m_2, B_\mu$, and $\lambda_0$ can be replaced by another more phenomenologically accessible set. That is, the masses of the states $h_0$ and $A_0$, the angle $\beta$ and the VEV $v$~\cite{Kalyniak}:

\begin{eqnarray}\label{parametros}
B_\mu &=&\frac{1}{2}(\lambda_0  v^{2} + m^{2}_{A_{0}}  )\, \text{sin}\, 2\beta,\\
\lambda_0 &=& \frac{m^{2}_{h_{0}}}{v^{2}}\Big(\frac{  m^{2}_{h_{0}}- m^{2}_{A_{0}} }{m^{2}_{h_{0}}-m^{2}_{A_{0}} \text{sin}^{2}\, 2\beta }\Big),\\
\text{tan}\, \alpha &=& \frac{ B_\mu \text{cot}\, 2\beta+ \sqrt{(B^{2}_\mu/\text{sin}^{2}\, 2\beta)-2\lambda_0 B_\mu v^{2} \text{sin}\, 2\beta+ \lambda^{2}_{0} v^{4}\text{sin}^{2}\, 2\beta  }  }{B_\mu -\lambda_0 v^{2} \text{sin}\, 2\beta},\label{alpha}   \\
m^{2}_{H_{0}} &=& \frac{B_\mu}{\text{sin}\, 2\beta}+ \sqrt{\frac{B^{2}_{\mu}}{\text{sin}^{2}\, 2\beta} -2\lambda_0 B_\mu v^{2} \text{sin}\, 2\beta +\lambda^{2}_{0} v^{4} \text{sin}^{2}\, 2\beta  }, \label{mH0}\\
m^{2}_{\sigma}&=&(\lambda_{56} + \lambda_{65})f^{2}=2\lambda_0 f^{2} \text{K}_\sigma, \label{masaescalar} \\
 1 < &\text{tan} & \ \beta  <  \sqrt{ \frac{2+2 \sqrt{\big(1-\frac{m^{2}_{h_0} }{m^{2}_{A_0}} \big) \big(1-\frac{m^{2}_{h_0} }{4 \pi v^{2}}\big) } }{ \frac{m^{2}_{h_0}}{m^{2}_{A_0}} \big(1+ \frac{m^{2}_{A_0}- m^{2}_{h_0}}{4 \pi v^{2}}  \big) } -1 }.
\end{eqnarray}

\noindent The variables $\lambda_{56}$ and $\lambda_{65}$ in Eq.~(\ref{masaescalar}) represent the coefficients of the quartic potential defined
in~\cite{JHEP09-2010}, both variables take values different from zero to achieve the collective breaking of the symmetry
and generate a quartic coupling of the Higgs boson~\cite{JHEP09-2010,Kalyniak}.

\subsection{The gauge sector}

The kinetic terms of the gauge fields in the BLHM are given as follows:

\begin{equation}\label{Lcinetico}
\mathcal{L}=\frac{f^{2}}{8} \text{Tr}(D_{\mu} \Sigma^{\dagger} D^{\mu} \Sigma) + \frac{F^{2}}{4} \text{Tr}(D_\mu \Delta^{\dagger} D^{\mu} \Delta),
\end{equation}

\noindent where the covariant derivatives are given by

\begin{eqnarray}\label{derivadasC}
D_{\mu}\Sigma&=&\partial_{\mu} \Sigma +i g_A A^{a}_{1\mu} T^{a}_L \Sigma- i g_B \Sigma A^{a}_{2\mu} T^{a}_L+ i g_{Y} B^{3}_{\mu}(T^{3}_{R}\Sigma-\Sigma T^{3}_{R}),\\
D_{\mu}\Delta&=&\partial_{\mu} \Delta +i g_A A^{a}_{1\mu} \frac{\tau^{a}}{2}  \Delta- i g_B \Delta A^{a}_{2\mu} \frac{\tau^{a}}{2}.
\end{eqnarray}

\noindent $T^{a}_{L}$ are the generators of the group $SO(6)_A$ corresponding to the subgroup $SU(2)_{LA}$, while $T^3_R$ represents
the third component of the $SO(6)_B$ generators corresponding to the $SU(2)_{LB} $ subgroup, these matrices are provided in~\cite{JHEP09-2010}.
$g_A$ and $A^{a}_{1\mu}$ denote the gauge coupling and field associated with the gauge bosons of $SU(2)_{LA}$. $g_B$ and $A^{a}_{2\mu}$
represent the gauge coupling and the field associated with $SU(2)_{LB}$, while $g_Y$ and $B^{3}_{\mu}$ denote the hypercharge and the field.
When $\Sigma$ and $\Delta$ get their VEVs, the gauge fields $A^{a}_{1\mu}$ and $A^{a}_{2\mu}$ are mixed to form a massless triplet
$A^{a}_{0\mu}$ and a massive triplet $A^{a}_{H\mu}$,

\begin{equation}\label{AA}
A^{a}_{0\mu}=\text{cos}\, \theta_g A^{a}_{1\mu} + \text{sin}\, \theta_g A^{a}_{2\mu}, \hspace{5mm} A^{a}_{H\mu}= \text{sin}\, \theta_g A^{a}_{1\mu}- \text{cos}\, \theta_g A^{a}_{2\mu},
\end{equation}

\noindent with the mixing angle

\begin{equation}\label{gagb}
s_g\equiv \sin \theta_g=\frac{g_A}{\sqrt{g_{A}^{2}+g_{B}^{2}} },\ \ c_g \equiv \cos \theta_g=\frac{g_B}{\sqrt{g_{A}^{2}+g_{B}^{2}} },
\end{equation}

\noindent
which are related to the electroweak gauge coupling $g$ through

\begin{equation}\label{g}
\frac{1}{g^{2}}=\frac{1}{g^{2}_A}+\frac{1}{g^{2}_B}.
\end{equation}

\noindent  On the other hand, the weak mixing angle is defined as

\begin{eqnarray}\label{angulodebil}
s_W&&\equiv\sin \theta_W = \frac{g_Y}{\sqrt{g^2+ g^{2}_Y }}, \\
c_W&&\equiv\cos \theta_W= \frac{g}{\sqrt{g^2+ g^{2}_Y }}.
\end{eqnarray}

\subsection{The Yang-Mills sector}

The gauge boson self-interactions arise from the following Lagrangian terms:

\begin{eqnarray}\label{YM}
\mathcal{L}=F_{1\mu \nu} F^{\mu \nu}_{1}  + F_{2\mu \nu} F^{\mu \nu}_{2},
\end{eqnarray}

\noindent
where $ F^{\mu \nu}_{1,2}$ are given by
\begin{eqnarray}
F^{\mu \nu}_{1} &=& \partial^{\mu} A^{\alpha \nu}_{1} - \partial^{\nu} A^{\alpha \mu}_{1} + g_{A} \sum_{b} \sum_{c} \epsilon^{a b c} A^{b \mu}_{1} A^{c \nu}_{1}, \\
F^{\mu \nu}_{2} &=& \partial^{\mu} A^{\alpha \nu}_{2} - \partial^{\nu} A^{\alpha \mu}_{2} + g_{B} \sum_{b} \sum_{c} \epsilon^{a b c} A^{b \mu}_{2} A^{c \nu}_{2}.
\end{eqnarray}

\noindent
In these equations, the indices $a$, $b$, and $c$ run over the three gauge fields~\cite{Martin:2012kqb}; $\epsilon^{a b c}$ is the anti-symmetric tensor.

\subsection{The fermion sector} \label{subsecfermion}

To construct the Yukawa interactions in the BLHM, the fermions must be transformed under the group $SO(6)_A$ or $SO(6)_B$.
In this model, the fermion sector is divided into two parts. First, the sector of massive fermions is represented by Eq.~(\ref{Ltop}).
This sector includes the top and bottom quarks of the SM and a series of new heavy quarks arranged in four multiplets, $Q$ and $Q^{\prime}$,
which transform under $SO(6)_A$, while $U^c$ and $U^{'c}_5$  are transformed under the group $SO(6)_B$. Second, the sector of
light fermions contained in  Eq.~(\ref{Lligeros}), in this expression, all the interactions of the remaining fermions of the SM with the
exotic particles of the BLHM are generated.

For massive fermions, the Lagrangian that describes them is given by~\cite{JHEP09-2010}
\begin{equation}\label{Ltop}
\mathcal{L}_t=y_1 f Q^{T} S \Sigma S U^{c} + y_2 f Q'^{T} \Sigma U^{c} +y_3 f Q^{T} \Sigma U'^{c}_{5} +y_b f q_{3}^{T}(-2 i T^{2}_{R} \Sigma) U^{c}_{b}+ h.c.,
\end{equation}

\noindent  where $ S = \text{diag} (1,1,1,1, -1, -1) $. The explicit representation of the multiplets involved in Eq.~(\ref{Ltop}) is provided in Refs.~\cite{JHEP09-2010,PhenomenologyBLH}. For simplicity, the Yukawa couplings are assumed to be real $y_1, y_2, y_3$ $\in R$.

For light fermions, the corresponding Lagrangian is~\cite{JHEP09-2010,PhenomenologyBLH,Martin:2012kqb}
\begin{equation}\label{Lligeros}
\mathcal{L}_{light}= \sum_{i=1,2} y_u f q^{T}_i \Sigma u^{c}_{i} + \sum_{i=1,2} y_{d} f q^{T}_{i}(-2i T^{2}_{R} \Sigma) d^{c}_i
+\sum_{i=1,2,3} y_e f l^{T}_i (-2i T^{2}_{R} \Sigma) e^{c}_i + h.c.
\end{equation}

\subsection{The currents sector}

The Lagrangian that describes the interactions of fermions with the gauge bosons is~\cite{JHEP09-2010,PhenomenologyBLH}
\begin{eqnarray}\label{LbaseW}
 \mathcal{L} &=& \bar{Q} \bar{\tau}^{\mu} D_{\mu}Q + \bar{Q}' \bar{\tau}^{\mu} D_{\mu}Q'- U^{c\dagger} \tau^{\mu} D_{\mu}U^{c}-  U'^{c\dagger} \tau^{\mu} D_{\mu}U'^{c} -  U_{b}^{c\dagger} \tau^{\mu} D_{\mu}U_{b}^{c} +\sum_{i=1,2}  q^{\dagger}_i \tau^{\mu} D_{\mu} q_i   \nonumber \\
 &+& \sum_{i=1,2,3}  l^{\dagger}_i \tau^{\mu} D_{\mu} l_i
 - \sum_{i=1,2,3}  e_i^{c\dagger} \tau^{\mu} D_{\mu} e^{c}_i - \sum_{i=1,2}  u_{i}^{c\dagger} \tau^{\mu} D_{\mu} u^{c}_{i} - \sum_{i=1,2}  d_{i}^{c\dagger} \tau^{\mu} D_{\mu} d^{c}_i,
\end{eqnarray}

\noindent  where $\tau^{\mu}$ and $\bar{\tau}^{\mu}$ are defined according to~\cite{Spremier}. On the other hand, the respective covariant derivatives are provided in Refs.~\cite{Cruz-Albaro:2022lks,Cruz-Albaro:2022kty}.

\section{Heavy Higgs boson decays at the BLHM} \label{H0decay}


Recently, the first experimental evidence for the decay of the Higgs boson of the SM ($ h^{SM}_0 $) into a photon and a $Z$ boson was presented, with a statistical significance of 3.4 standard deviations~\cite{ATLAS:2020qcv,CMS:2023}. The result is derived from a combined analysis of the searches performed by the ATLAS and CMS collaborations. In addition to this search channel, the diphoton decay of the SM Higgs is also a verified fact~\cite{CMS:2020xrn}. The $ h^{SM}_0 \to \gamma \gamma, \gamma Z$ processes are essential particle physics as they are sensitive to possible contributions from physics beyond the SM and can even probe scenarios where the SM-like Higgs boson emerges. Although the $ h^{SM}_0 \to \gamma \gamma, \gamma Z$ decay channels have small branching fractions, they provide a clean final state topology which can reconstruct the diphoton invariant mass and photon-$Z(\to $\textit{ll}, $l=e\ \text{or}\ \mu $) invariant mass with high precision~\cite{ATLAS:2020qcv,CMS:2020xrn}.
A new window to explore physics not described by the SM has opened with the discovery of the first fundamental scalar particle $ h^{SM}_0 $. Many of the extended models postulate the existence of heavy states. Linked to this, the search for additional scalar particles is being carried out by the ATLAS and CMS collaborations at the CERN LHC using data from increasingly energetic collisions. These facts motivate our study of the production of the $ H_0 $ Higgs of the BLHM at the LHC and the FCC-hh.
For this purpose, we calculate the total decay width of the $H_0$  Higgs, where we will consider direct search channels with SM final states. In the following, we describe the different decay modes of the  Higgs boson $ H_0 $.

\subsection{Two-body decays  of the  Higgs boson $ H_0 $ at tree level }

\begin{figure}[H]
\begin{center}
\includegraphics[width=8.0cm,height=4.5cm]{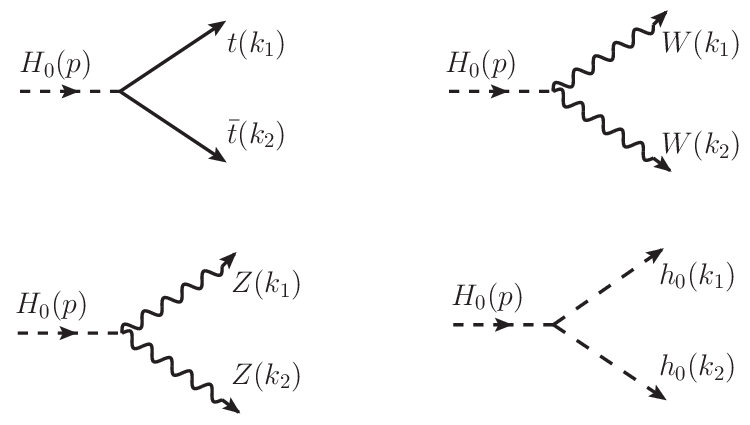}
\caption{\label{tree1} Feynman diagrams corresponding to the two-body decays of the $ H_0 $ Higgs at tree level.}
\end{center}
\end{figure}

In the BLHM, the Feynman diagrams representing the two-body decays of the $ H_0 $ Higgs at tree level are shown in Fig.~\ref{tree1}.
To calculate the partial decay widths of $H_0$, we use the Feynman rules for the interaction vertices provided in  Refs.~\cite{Cruz-Albaro:2023pah,Cruz-Albaro:2022lks,Cruz-Albaro:2022kty,Gutierrez-Rodriguez:2023sxg,Aranda:2021kza}.

\noindent  The decay widths of $ H_0 \to \bar{t}t,W W, ZZ, h_0 h_0 $ can be written as follows

\begin{eqnarray}
\Gamma(H_0\to \bar{t}t)=\frac{N_c g_{H_0 tt}^2 m_{H_0}}{8\pi} \left(1-\frac{4m_t^2}{m_{H_0}^2}\right)^{3/2},
\end{eqnarray}

\begin{eqnarray}
\Gamma(H_0\to WW)=\frac{g_{H_0 WW}^2 m_{H_0}^3}{64 \pi m_W^4}\sqrt{1-\frac{4m_W^2}{m_{H_0}^2}} \left( 1-\frac{4m_W^2}{m_{H_0}^2} + \frac{12m_W^4}{m_{H_0}^4} \right),
\end{eqnarray}

\begin{eqnarray}
\Gamma(H_0\to ZZ)=\frac{g_{H_0 ZZ}^2 m_{H_0}^3}{128 \pi m_Z^4}\sqrt{1-\frac{4m_Z^2}{m_{H_0}^2}} \left( 1 -\frac{4m_Z^2}{m_{H_0}^2} + \frac{12m_Z^4}{m_{H_0}^4} \right),
\end{eqnarray}

\begin{eqnarray}
\Gamma(H_0\to h_{0}h_{0})=\frac{ g_{H_0 h_{0}h_{0}}^2}{32\pi m_{H_0}} \sqrt{1-\frac{4m_{h_{0}}^2}{m_{H_0}^2}},
\end{eqnarray}

\noindent where $ N_c =3$, is the color factor and $g_{H_0 tt}, g_{H_0 WW}, g_{H_0 ZZ}$ and $g_{H_0 h_{0}h_{0}}$  represent the couplings of the interaction vertices involved in the tree-level processes, which are given in Refs.~\cite{Cruz-Albaro:2023pah,Cruz-Albaro:2022lks,Cruz-Albaro:2022kty,Gutierrez-Rodriguez:2023sxg,Aranda:2021kza}.

\subsection{Three-body decays of the Higgs boson $ H_0 $ at tree level}

\begin{figure}[H]
\begin{center}
\includegraphics[width=13.5cm,height=8.5cm]{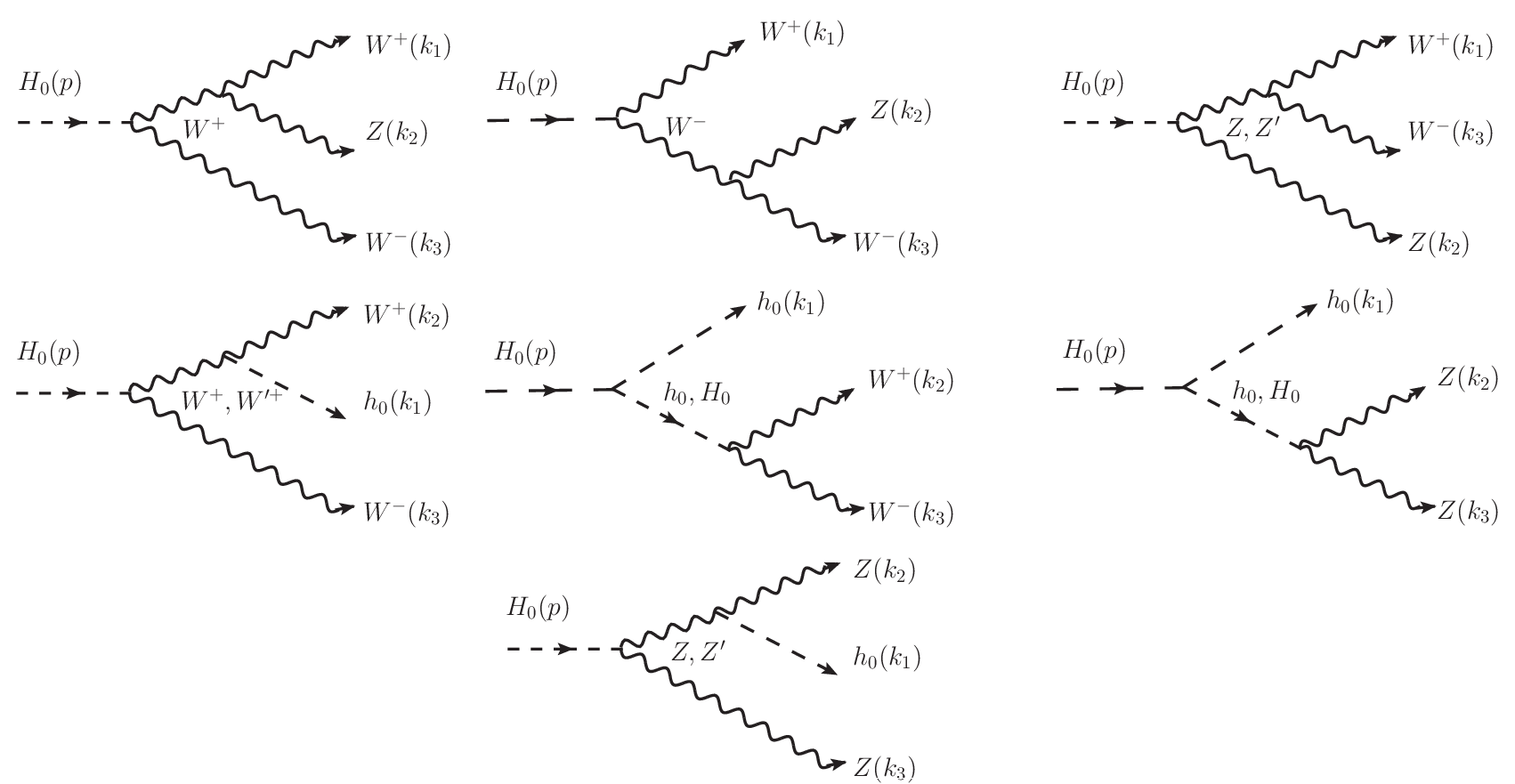}
\caption{\label{tree2} Feynman diagrams corresponding to the three-body decays of the $ H_0 $ Higgs at tree level.}
\end{center}
\end{figure}

Concerning the three-body decays of the $ H_0 $ Higgs, the Feynman diagrams that arise for these processes are shown in Fig.~\ref{tree2}. The scalar particles and gauge bosons of the SM and the BLHM mediate these processes.
The Feynman rules for the interaction vertices are given in  Refs.~\cite{Cruz-Albaro:2023pah,Cruz-Albaro:2022lks,Cruz-Albaro:2022kty,Gutierrez-Rodriguez:2023sxg}.
We only provide the decay amplitudes because the expressions generated for the partial widths are quite lengthy.  The analytical expressions for the decay widths of the $H_0  $ Higgs decaying to three bodies can be calculated using the generic formula described in Eq.~(\ref{width3})~\cite{pdg:2023,Barradas:1996xb},

\begin{eqnarray} \label{width3}
\frac{d\Gamma(H_0 \to ABC)}{dx_a dx_b}=\frac{m_{H_0}}{256\pi^3}\vert \mathcal{M}(H_0 \to ABC) \vert^2.
\end{eqnarray}

It is worth mentioning that for some decay channels of the $H_0$ Higgs, specific interaction vertices ($Z^\prime WW$, $h_0 WW^\prime$, and $h_0 ZZ^\prime$) cancel out.
 This happens because in the BLHM the gauge couplings $g_{A}$ and $g_{B}$, associated to the gauge bosons $SU(2)_{LA}$ and $SU(2)_{LB}$, can be parametrized in a more phenomenological fashion in terms of a mixing angle $\theta_{g}$, $\tan \theta_{g}=g_{A}/g_{B}$.
For simplicity, it is assumed that $\tan \, \theta_{g}=1$~\cite{Cruz-Albaro:2022lks,PhenomenologyBLH}, which implies that the gauge couplings $g_{A}$ and $g_{B}$ are equal, i.e.,  $  \sin \theta_g = \cos \theta_g $ ($s_g=c_g$).
This fact leads to no contribution from specific decay amplitudes. The only contributing amplitudes are given in Eqs.~(\ref{m3})-(\ref{m1}).

\begin{eqnarray} \label{m3}
\mathcal{M}(H_0 \to WZZ)=&ig_{WWZ}\bigg[-\frac{g_{H_{0}WW}}{(k_1+k_2)^2-m_W^2}\bigg((k_1^\mu+2k_2^\mu)g^{\alpha\nu}-(2k_1^\nu+k_2^\nu)g^{\alpha\mu}\nonumber\\
&+\big(-2k_2^\alpha+\frac{m_Z^2}{m_W^2}(k_1^\alpha+k_2^{\alpha})\big)g^{\mu\nu}+\frac{(k_1^\alpha+k_2^\alpha)\big(k_1^\mu k_1^\nu-k_2^\mu k_2^\nu\big)}{m_W^2}\bigg)\nonumber\\
&+\frac{g_{H_{0}WW}}{(k_2+k_3)^2-m_W^2}\bigg((2k_2^\alpha+k_3^\alpha)g^{\mu\nu}+\frac{(k_3^\alpha k_3^\nu-k_2^\alpha k_2^\nu)(k_2^\mu+k_3^\mu)}{m_W^2}\nonumber\\
&+\big(-2k_2^\mu+\frac{m_Z^2}{m_W^2}(k_2^\mu+k_3^\mu) \big)g^{\alpha \nu}-(k_2^\nu+2k_3^\nu)g^{\alpha \mu}\bigg)\nonumber\\
&+\frac{g_{H_{0}ZZ}}{(k_1+k_3)^2-m_Z^2}\bigg(-(k_1^\mu+2k_3^\mu)g^{\alpha\nu}-(k_1^\nu-k_3^\nu)g^{\alpha\mu}\nonumber\\
&+(2k_1^\alpha+k_3^\alpha)g^{\mu\nu}+\frac{(k_3^\alpha k_3^\mu-k_1^\alpha k_1^\mu)(k_1^\nu+k_3^\nu)}{m_Z^2}\bigg)\bigg]\epsilon_\mu^*(k_1)\,\epsilon_\nu^*(k_2)\epsilon_\alpha^*(k_3),
\end{eqnarray}

\begin{eqnarray} \label{m2}
\mathcal{M}(H_0 \to h_0WW)=&i \bigg[\bigg(\frac{g_{h_0 H_{0}H_{0}}\,g_{H_{0}WW}}{(k_2+k_3)^2-m_{H_{0}}^2}+\frac{g_{H_{0}h_0h_0}\,g_{h_0 WW}}{(k_2+k_3)^2-m_{h_0}^2}+\frac{g_{H_{0}WW}\,g_{{h_0}WW}}{(k_1+k_2)^2-m_W^2}\bigg)g^{\mu\nu}\nonumber\\
&-\frac{g_{h_0 WW}\,g_{H_{0}WW} \big(k_1^\mu+k_2^\mu\big)\big(k_1^\nu+k_2^\nu\big)}{m_W^2\big((k_1+k_2)^2-m_W^2\big)}\bigg]\epsilon_\mu^*(k_2)\,\epsilon_\nu^*(k_3),
\end{eqnarray}

\begin{eqnarray} \label{m1}
\mathcal{M}(H_{0} \to h_{0} ZZ)=&i \bigg[\bigg(\frac{g_{h_{0} H_{0}H_{0}}\,g_{H_{0}ZZ}}{(k_2+k_3)^2-m_{H_{0}}^2}+\frac{g_{H_{0}h_{0} h_{0} }\,g_{h_{0} ZZ}}{(k_2+k_3)^2-m_{h_{0}}^2}+\frac{g_{H_{0}ZZ}\,g_{{h_{0} }ZZ}}{(k_1+k_2)^2-m_Z^2}\bigg)g^{\mu\nu}\nonumber\\
&-\frac{g_{h_{0} ZZ}\,g_{H_{0}ZZ} \big(k_1^\mu+k_2^\mu\big)\big(k_1^\nu+k_2^\nu\big)}{m_Z^2\big((k_1+k_2)^2-m_Z^2\big)}\bigg]\epsilon_\mu^*(k_2)\,\epsilon_\nu^*(k_3).
\end{eqnarray}

\subsection{Two-body decays  of the  Higgs boson $ H_0 $ at one-loop level}

\begin{figure}[H]
\begin{center}
\includegraphics[width=15.5cm,height=7.5cm]{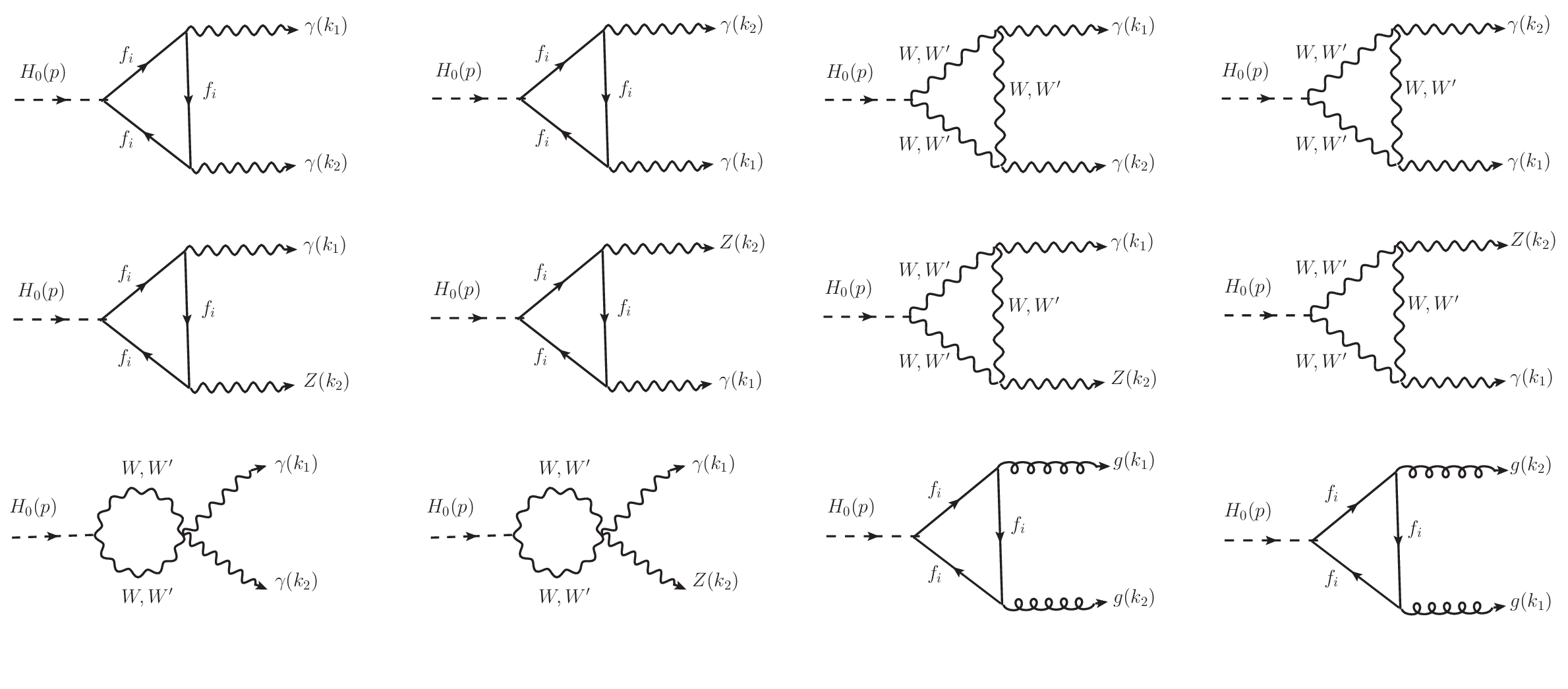}
\caption{\label{DFloop} Feynman diagrams corresponding to the two-body decays of the $ H_0 $ Higgs at one-loop level. The notation $f_i$ represents the $ t, T, T^{2/3}, T_5$, $ T_6 $ quarks.}
\end{center}
\end{figure}

This subsection determines the amplitudes and partial decay widths of the $H_0$ Higgs at the one-loop level. In Fig.~\ref{DFloop}, we show the Feynman diagrams associated with the $H_0 \to  \gamma\gamma, \gamma Z, gg$ decays which are mediated by fermions ($ t; T; T^{2/3}; T_5; T_6 $) and gauge bosons ($ W^{\pm}; W'^{\pm} $) of the SM and BLHM.
These $H_0$ decays are absent at the tree level in the BLHM. However, they arise at a one-loop level, which is of great interest since they not only help to examine higher-order corrections to the theory but also provide information about possible contributions from new particles circulating in the loop.

In the following, all calculations at the one-loop level will be carried out using the unitary gauge and the Passarino-Veltman (PV) reduction scheme~\cite{Denner:2005nn}. In this way, we find that the total amplitude for the first decay $H_0\to \gamma \gamma$ can be written as

\begin{eqnarray} \label{M2photon}
\mathcal{M}(H_0 \to \gamma \gamma)= (A_f^{\gamma \gamma}+A_W^{\gamma \gamma}+A_{W^\prime}^{\gamma \gamma}) \big(k_1^\nu k_2^\mu-k_1.k_2 g^{\mu\nu}\big) \epsilon^{*}_{\mu}(k_1)\epsilon^{*}_{\nu}(k_2),
\end{eqnarray}

\noindent where the form factors $ A_f^{\gamma \gamma}$, $A_W^{\gamma \gamma}$, $A_{W^\prime}^{\gamma \gamma} $ are defined as follows

\begin{footnotesize}
\begin{eqnarray}
A_f^{\gamma\gamma}&=&\frac{N_c}{2\pi^2m_{H_0}^2}\sum_{f=t,T,T_5,T_6,T^{2/3}} \:g_{H_0ff}(g_{Aff})^2m_f Q_f\bigg(2-(m_{H_0}^2-4m_f^2) C_{0}(m_{H_0}^2,0,0,m_f^{2},m_f^{2},m_f^{2})\bigg),
\end{eqnarray}
\end{footnotesize}

\begin{eqnarray}
A_W^{\gamma\gamma}=\frac{g_{H_0WW}(g_{WWA})^2}{8\pi^2m_{H_0}^2}\bigg(6(m_{H_0}^2-2m_W^2) C_{0}(m_{H_0}^2,0,0,m_W^{2},m_W^{2},m_W^{2}) -\frac{m_{H_0}^2}{m_W^2}-6\bigg),
\end{eqnarray}

\begin{eqnarray}
A_{W^{\prime}}^{\gamma \gamma}=\frac{g_{H_0 W^{\prime} W^{\prime}}(g_{W^{\prime}W^{\prime}A})^2}{8\pi^2m_{H_0}^2}\bigg(6(m_{H_0}^2-2m_{W^{\prime}}^2) C_{0}(m_{H_0}^2,0,0,m_{W^{\prime}}^{2},m_{W^{\prime}}^{2},m_{W^{\prime}}^{2}) -\frac{m_{H_0}^2}{m_{W^{\prime}}^2}-6\bigg).
\end{eqnarray}

\noindent The labels $ f $ and $ W $ ($ W^{\prime} $) of the form factors $ A_f^{\gamma\gamma} $ and $ A_W^{\gamma\gamma} $ ($ A_{W^{\prime}}^{\gamma\gamma} $), respectively, represent the contributions generated due to the quarks and gauge bosons circulating in the loop of the diphoton decay of $ H_0 $ Higgs. On the other hand, $ C_{0}(m_{H_0}^2,0,0,m_f^{2},m_f^{2},m_f^{2}) $, $ C_{0}(m_{H_0}^2,0,0,m_W^{2},m_W^{2},m_W^{2}) $ and $ C_{0}(m_{H_0}^2,0,0,m_{W^{\prime}}^{2},m_{W^{\prime}}^{2},m_{W^{\prime}}^{2}) $ denote the scalar functions of PV.

Using Eq.~(\ref{M2photon}), we find that the decay width for the $H_0\to \gamma \gamma$ process is

\begin{eqnarray}
\Gamma(H_0 \to \gamma\gamma)=\frac{1}{64\pi} \mid A^{\gamma\gamma}_f+A^{\gamma\gamma}_W +A^{\gamma\gamma}_{W'} \mid^2  m_{H_0}^3.
\end{eqnarray}

The one-loop decay amplitude for the second process $ H_{0} \to \gamma Z $ is

\begin{eqnarray}
\mathcal{M}(H_0 \to \gamma Z)= (A_f^{\gamma Z}+A_W^{\gamma Z} +A_{W^{\prime}}^{\gamma Z}) \big(k_1^\nu k_2^\mu-k_1.k_2 g^{\mu\nu}\big) \epsilon^{*}_{\mu}(k_1)\epsilon^{*}_{\nu}(k_2),
\end{eqnarray}

\noindent with

\begin{footnotesize}
\begin{eqnarray}
A_f^{\gamma Z}&=&\frac{N_c}{2\pi^2(m_{H_0}^2-m_Z^2)^2}\sum_{f=t,T,T_5,T_6,T^{2/3}}m_f Q_f\bigg[ \:g_{H_0ff}\,g_{Aff}\,g_{Zff}\bigg((m_{H_0}^2-m_Z^2)(m_{Z}^2+4m_f^2-m_{H_0}^2) \nonumber\\ &\times& C_{0}(m_{H_0}^2,0,0,m_f^{2},m_f^{2},m_f^{2})+2\bigg)-2m_Z^2\bigg(B_0(m_Z^2,m_f^2,m_f^2)-B_0(m_{H_0}^2,m_f^2,m_f^2)\bigg)\bigg],
\end{eqnarray}
\end{footnotesize}

\begin{eqnarray}
A_W^{\gamma Z}=&\frac{g_{H_0WW}\,g_{WWA}\,g_{WWZ}}{16\pi^2t_W^4(1-t_Z^2)m_{H_0}^2}\bigg[\bigg(12t_W^4-2t_W^2(t_Z^2-1)-t_Z^2\bigg)\nonumber\\
&\times \bigg(t_Z^2\big(B_{0}(m_{H_0}^2,m_W^2,m_W^2)-B_{0}(m_Z^2,m_W^2,m_W^2)-1\big)+1\bigg)\nonumber\\
&-2t_W^2(t_Z^2-1)\bigg(12t_W^4+6t_W^2(t_Z^2-1)-t_Z^2(2t_Z^2-1)\bigg)\nonumber\\
&\times m_{H_0}^2C_{0}(m_{H_0}^2,m_Z^2,0,m_W^2,m_W^2,m_W^2)\bigg],
\end{eqnarray}

\noindent and

\begin{eqnarray}
A_{W^{\prime}}^{\gamma Z}=&\frac{g_{H_0W^{\prime}W^{\prime}}\,g_{W^{\prime}W^{\prime}A}\,g_{W^{\prime}W^{\prime}Z}}{16\pi^2t_{W^{\prime}}^4(1-t_Z^2)m_{H_0}^2}\bigg[\bigg(12t_{W^{\prime}}^4-2t_{W^{\prime}}^2(t_Z^2-1)-t_Z^2\bigg)\nonumber\\
&\times \bigg(t_Z^2\big(B_{0}(m_{H_0}^2,m_{W^{\prime}}^2,m_{W^{\prime}}^2)-B_{0}(m_Z^2,m_{W^{\prime}}^2,m_{W^{\prime}}^2)-1\big)+1\bigg)\nonumber\\
&-2t_{W^{\prime}}^2(t_Z^2-1)\bigg(12t_{W^{\prime}}^4+6t_{W^{\prime}}^2(t_Z^2-1)-t_Z^2(2t_Z^2-1)\bigg)\nonumber\\
&\times m_{H_0}^2C_{0}(m_{H_0}^2,m_Z^2,0,m_{W^{\prime}}^2,m_{W^{\prime}}^2,m_{W^{\prime}}^2)\bigg],
\end{eqnarray}

\noindent being $t_W=\frac{m_W}{m_{H_0}}$, $t_Z=\frac{m_Z}{m_{H_0}}$ and $t_{W^{\prime}}=\frac{m_{W^{\prime}}}{m_{H_0}}$.
It is worth mentioning that the form factors $ A_f^{\gamma Z} $, $ A_W^{\gamma Z} $ and $ A_{W^{\prime}}^{\gamma Z} $  provide finite results, i.e. these are free of ultraviolet divergences.
For the $ H_{0} \to \gamma Z $ decay, their corresponding decay width is

\begin{eqnarray}
\Gamma(H_0\to \gamma Z)=\frac{1}{32\,\pi\,m_{H_0}^3} \mid A_f^{\gamma Z}+A_W^{\gamma Z} +A_{W^{\prime}}^{\gamma Z} \mid^2 (m_{H_0}^2-m_Z^2)^3.
\end{eqnarray}

Finally, the decay amplitude of the  process $ H_0\to gg$ is as follows

\begin{eqnarray} \label{Mgg}
\mathcal{M}(H_0 \to gg)= A^{gg} \big(k_1^\nu k_2^\mu-k_1.k_2 g^{\mu\nu}\big) \epsilon^{*a}_{\mu}(k_1)\epsilon^{*b}_{\nu}(k_2)\delta_{ab},
\end{eqnarray}

\noindent where

\begin{footnotesize}
\begin{eqnarray}
A^{gg}=\frac{N_c}{4\pi^2m_{H_0}^2}\sum_{f=t,T,T_5,T_6,T^{2/3}} \:g_{H_0ff}(g_{g_sff})^2m_f\bigg(2-(m_{H_0}^2-4m_f^2) C_{0}(m_{H_0}^2,0,0,m_f^{2},m_f^{2},m_f^{2})\bigg).
\end{eqnarray}
\end{footnotesize}

\noindent The corresponding decay width for the $H_0 \to gg$ process is given by:

\begin{eqnarray}
\Gamma(H_0 \to gg)=\frac{1}{8\pi} \mid A^{gg} \mid^2  m_{H_0}^3.
\end{eqnarray}

\section{Numerical results} \label{results}

To carry out our numerical analysis of the $ H_0 $ Higgs production in the context of the BLHM, we briefly review some free parameters of the model of interest and provide in Table~\ref{parametervalues} the values assigned to the remaining free parameters.

As discussed in Refs.~\cite{Cruz-Albaro:2023pah,Cruz-Albaro:2022kty}, the three Yukawa couplings, $y_1$, $y_2$, and $y_3$ generate two study scenarios in the BLHM:

\begin{itemize}
\item Scenario {\bf a} ($y_2 < y_3$), $y_1=0.61$, $y_2=0.35$ and $y_3=0.84$~\cite{Cruz-Albaro:2023pah,Cruz-Albaro:2022kty,Cruz-Albaro:2022lks},
\item Scenario  {\bf b} ($y_2 > y_3$), $y_1=0.61$, $y_2=0.84$ and $y_3=0.35$~\cite{Cruz-Albaro:2023pah,Cruz-Albaro:2022kty,Cruz-Albaro:2022lks}.
\end{itemize}

\noindent The restrictions on the Yukawa couplings arise due to perturbativity requirements that set an upper bound of $ 4 \pi $ on $y_i, i=1,2,3$~\cite{Altmannshofer:2010zt}.
On the other hand, the BLHM is designed to overcome the fine-tuning problem, which suggests that the masses of the new quarks must be less than about 2 TeV~\cite{JHEP09-2010}. The expressions for the masses of the new heavy quarks involved in our calculations are, in turn, directly related to the Yukawa couplings $y_i$.
 In this way, scenarios {\bf a} and {\bf b} mentioned above offer realistic values of the Yukawa couplings that evade the fine-tuning constraints and satisfy the perturbativity requirements.
In the following numerical analysis, our results are generated only for scenario $a$. Scenario $b$ provides nearly identical results.
The other input parameters involved in our analysis of the production of the $H_0$ Higgs are shown in Table~\ref{parametervalues}.

\begin{table}[H]
\caption{Values assigned to the free parameters involved in our numerical analysis at the BLHM.
\label{parametervalues}}
\centering
\begin{tabular}{|c | c | c |}
\hline
\hspace{0.5cm} $ \textbf{Parameter} $ \hspace{0.5cm}  &  \hspace{1.2cm}  $\textbf{Value} $ \hspace{1.2cm}  &  \hspace{0.5cm}   $ \textbf{Reference} $ \hspace{0.5cm} \\
\hline
\hline
$ m_{h_{0}}  $  &   $ 125.25\  \text{GeV} $ &  \cite{Workman:2022ynf}  \\
\hline
$ m_{H_{0}}  $  &   $ 1015\  \text{GeV} $ &   \cite{ATLAS:2020gxx}  \\
\hline
$ \tan \beta $  &    $ 3 $  &  \cite{Cruz-Albaro:2023pah,Cruz-Albaro:2022kty,Cruz-Albaro:2022lks} \\
\hline
$ g_{A}=g_{B} $  &   $ \sqrt{2}\, g $ & \cite{Cruz-Albaro:2023pah,Cruz-Albaro:2022kty,Cruz-Albaro:2022lks}  \\
\hline
$ f $  &  $ [1000, 3000]\   \text{GeV} $ &  \cite{JHEP09-2010,Cruz-Albaro:2023pah, Cruz-Albaro:2022kty,Cruz-Albaro:2022lks,Kalyniak}  \\
\hline
$ F $  &   $ [3000, 6000] \ \text{GeV} $ &  \cite{JHEP09-2010,Cruz-Albaro:2023pah,Cruz-Albaro:2022kty,Cruz-Albaro:2022lks,Kalyniak} \\
\hline
\end{tabular}
\end{table}

Due to the characteristics of the BLHM, this is based on two independent global symmetries that break into diagonal subgroups at different energy scales, $ f $ and $ F $. These scales represent the scales of the new physics. Therefore, it is convenient to analyze the $H_0$ production cross-section as a function of the energy scales $f$ and $F$ since the masses of the particles circulating in the loop of the $H_0 \to \gamma \gamma, \gamma Z, gg$ processes depend on the scales $ f $ and $F$.
On the other hand, one-loop decays of $H_0$ into $\gamma \gamma$, $\gamma Z$, and  $gg$ will be helpful to test the consistency of the current parameter space of the BLHM.

\begin{figure}[H]
\center
\subfloat[]{\includegraphics[width=8.10cm]{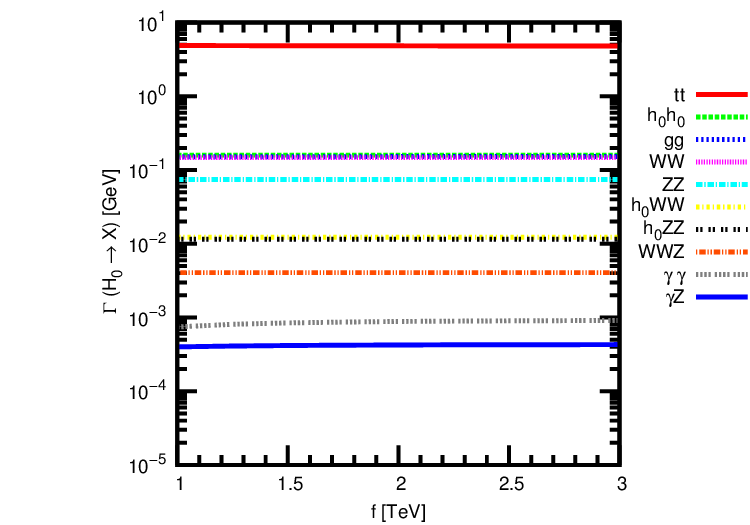}}
\subfloat[]{\includegraphics[width=8.10cm]{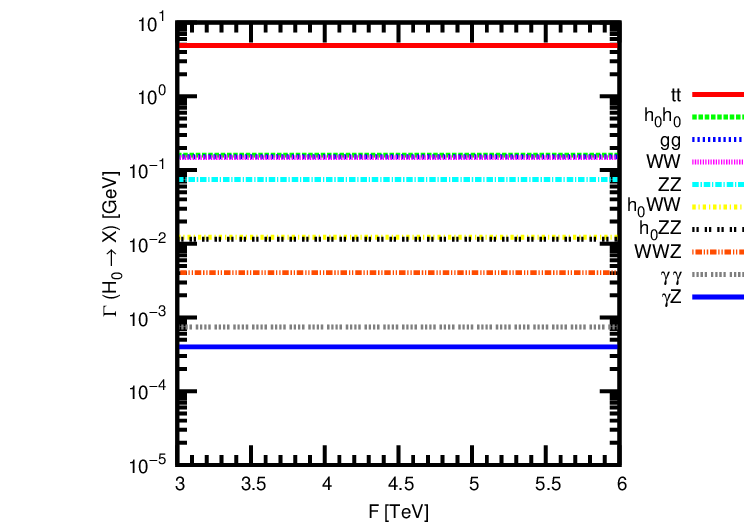}}
\caption{ \label{widhts} Decay widths for the $H_0\to X$ processes where $X =$ $tt$, $h_0h_0$, $gg$, $WW$, $ZZ$, $h_0WW$, $h_0ZZ$, $WWZ$, $\gamma\gamma$, $\gamma Z$. a) $\Gamma(H_0\to X)$ as a function of the $f$ energy scale (with the fixed value of $F=4000$ GeV). b) $\Gamma(H_0\to X)$ as a function of the $F$ energy scale (with the fixed value of $f=1000$ GeV).
}
\end{figure}

For the purposes mentioned above, we begin by presenting a numerical analysis of the decay widths for the  $H_0 \to tt, h_0h_0, gg, WW, ZZ, h_0WW, h_0ZZ, WWZ, \gamma\gamma, \gamma Z$ processes.
In this way, in Fig.~\ref{widhts} we show the behavior of $\Gamma(H_0\to X)$ vs. $f$ and $\Gamma(H_0\to X)$ vs. $F$, where $\Gamma(H_0\to X)$ denotes the partial decay width of the $ H_0 $ Higgs.  From the first plot (see Fig.~\ref{widhts}(a)), we can appreciate that the dominant and subdominant contributions to the decay width of $H_0$ ($ \Gamma_{H_0} $) over the whole analysis interval of scale $f$ come from the tree-level decays of $H_0$ into $\bar{t}t$ and $h_0 h_0$, respectively.
The numerical contributions of these decays are $\Gamma(H_0\to \bar{t}t)\sim 10^{0}$ GeV and $\Gamma(H_0\to h_0 h_0) \sim 10^{-1}$ GeV.
 For this last contribution, the decay channels $ H_0\to gg $ and $ H_0\to WW $ also contribute with the same order of magnitude although slightly smaller than $\Gamma(H_0\to h_0 h_0)$. Notice that the $ H_0\to gg $  process is a one-loop decay. Other decay modes that contribute in lower order but contribute significantly to $ \Gamma_{H_0} $ are the $H_0 \to ZZ, h_0 WW, h_0 ZZ, WWZ$ decays, whose associated decay widths are of $10^{-2}-10^{-3}$ GeV when $f \in [1000, 3000]$ GeV, which are all tree-level decays.
  For the remaining one-loop decays, we find that the $H_0 \to \gamma \gamma, \gamma Z$ decays provide suppressed contributions to the decay width of the $H_0$ Higgs: $\Gamma(H_0\to \gamma \gamma) \sim \Gamma(H_0\to \gamma Z) \sim 10^{-4}$ GeV.
Regarding the second plot (see Fig.~\ref{widhts}(b)) which examines the dependence of $\Gamma(H_0\to X)$ on the energy scale $F$, we observe that the largest contributions to $  \Gamma_{H_0} $ are again generated by the  $H_0 \to \bar{t}t, h_0 h_0$ decays: $\Gamma(H_0\to \bar{t}t)\sim 10^{0}$ GeV and $\Gamma(H_0\to h_0 h_0) \sim 10^{-1}$ GeV for the interval of $F=[3000,6000]$ GeV.
The processes $H_0\to gg, WW$ and $H_0\to ZZ, h_0 WW, h_0 ZZ$ also contribute considerably to the decay width of $H_0$, for these decays we find that the corresponding numerical estimates are $\Gamma(H_0\to gg)\sim \Gamma(H_0\to  WW)\sim 10^{-1}$ GeV and $\Gamma(H_0\to ZZ)\sim \Gamma(H_0\to  h_0 WW) \sim \Gamma(H_0\to h_0 ZZ)\sim 10^{-2}$ GeV.
 Finally, the curves that provide the most suppressed contributions compared to the main contribution come from the $H_0\to WWZ$ and $H_0\to \gamma \gamma, \gamma Z$ decays, their predicted numerical magnitudes are 3 and 4 orders of magnitude smaller than $\Gamma(H_0\to \bar{t}t)$, while the $F$ scale acquires values in the range of 3000 to 6000 GeV.
In summary, the numerical evaluation tells us that the tree-level decay channel $H_0 \to \bar{t}t$ provides the most significant contribution to the $H_0$ decay width. In contrast, the $H_0 \to \gamma Z$ decay at the one-loop level generates the most minor contribution.  On the other hand, we find that $\Gamma(H_0\to X)$ shows a greater sensitivity to the $f$ scale compared to another scale of the new physics, $F$.

\begin{figure}[H]
\center
\subfloat[]{\includegraphics[width=8.10cm]{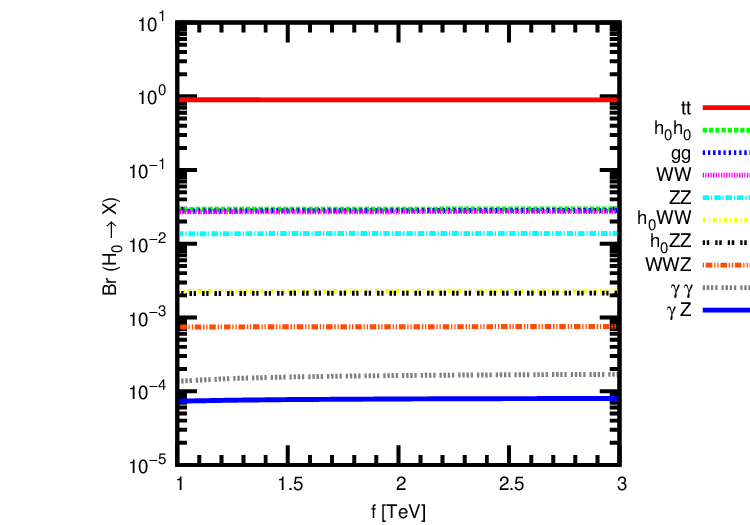}}
\subfloat[]{\includegraphics[width=8.10cm]{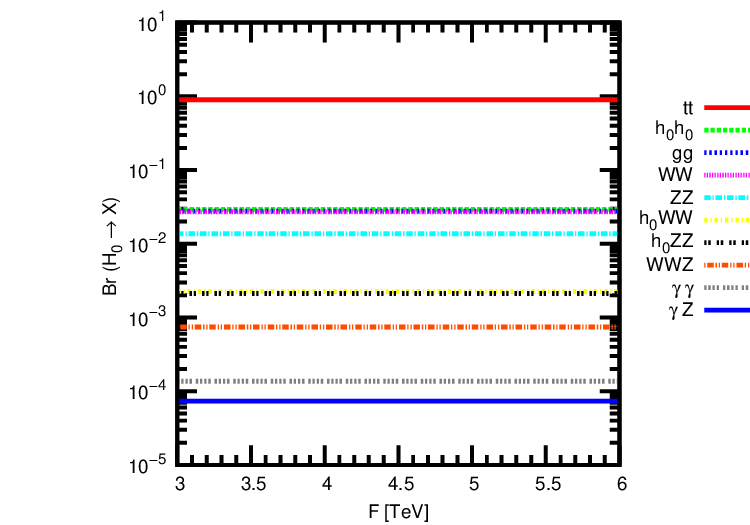}}
\caption{\label{branchings} The branching ratios for the $H_0\to X$ processes where $X =$ $tt$, $h_0h_0$, $gg$, $WW$, $ZZ$, $h_0WW$, $h_0ZZ$, $WWZ$, $\gamma\gamma$, $\gamma Z$. a) $\text{Br}(H_0\to X)$ as a function of the $f$ energy scale (with the fixed value of $F=4000$ GeV). b) $\text{Br}(H_0\to X)$ as a function of the $F$ energy scale (with the fixed value of $f=1000$ GeV).
}
\end{figure}

We also calculate the branching ratios of the Higgs boson $H_0$ as a function of the energy scale $f$ and $F$, as shown in  Fig.~\ref{branchings}.
The plots are obtained considering the total decay width of the $H_0$ Higgs, which contains the following decay modes:
 $\bar{t}t,WW, ZZ, h_0h_0, h_0WW, h_0ZZ, WWZ, \gamma\gamma, \gamma Z, gg$.
 From Fig.~\ref{branchings}(a), we can see the curves that represent the estimates of the branching ratios versus the $f$ scale when it takes values from 1000 to 3000 GeV. The $ H_0 \to \bar{t}t $ decay yields the highest contribution; its associated branching ratio is $ \text{Br}(H_0 \to \bar{t}t ) \sim 10^{-1} $.
 On the opposite side, we find that the most suppressed contribution is generated by the $ H_0 \to \gamma Z $ decay whose numerical magnitude of its branching ratio is  $ 10^{-5} $.
  The remaining branching ratios turn out to be $ \text{Br}(H_0 \to h_0 h_0 ) \sim \text{Br}(H_0 \to gg ) \sim \text{Br}(H_0 \to WW ) \sim \text{Br}(H_0 \to ZZ ) \sim 10^{-2} $,  $ \text{Br}(H_0 \to h_0  WW ) \sim \text{Br}(H_0 \to h_0 ZZ ) \sim 10^{-3} $ and $ \text{Br}(H_0 \to   WW Z ) \sim \text{Br}(H_0 \to \gamma \gamma) \sim 10^{-4} $ for $ f=[1000,3000] $ GeV.
 Concerning Fig.~\ref{branchings}(b), in this figure, we describe the behavior of $ \text{Br}(H_0 \to X )$ when the energy scale $F$ takes values within the set analysis interval. As can be appreciated in the corresponding plot, the dominant branching ratios correspond to the tree-level decays ($ H_0 \to \bar{t}t, h_0h_0$) of the $H_0$ Higgs while the minor contributions arise for one-loop decays ($H_0 \to \gamma \gamma, \gamma Z $):  $ \text{Br}(H_0 \to \bar{t}t ) = 8.97\times 10^{-1}$,  $ \text{Br}(H_0 \to h_0 h_0 ) = 2.92 \times 10^{-2}$,  $ \text{Br}(H_0 \to \gamma \gamma ) = 1.36 \times 10^{-4}$ and  $ \text{Br}(H_0 \to \gamma Z ) = 7.33 \times 10^{-5}$.
 For these cases, the numerical predictions of $\text{Br}(H_0 \to X)$ produce constant values over the whole range of $F$ analysis, this also happens for the rest of the $H_0$ decays, specifically $H_0 \to gg, WW, ZZ, h_0 WW, h_0 ZZ, WWZ$ which generate the following branching ratios:
  $ \text{Br}(H_0 \to gg ) = 2.78 \times 10^{-2}$,  $ \text{Br}(H_0 \to WW ) = 2.71 \times 10^{-2}$,  $ \text{Br}(H_0 \to Z Z ) = 1.37 \times 10^{-2}$,  $ \text{Br}(H_0 \to h_0 WW ) = 2.25\times 10^{-3}$,  $ \text{Br}(H_0 \to h_0 Z Z ) = 2.12 \times 10^{-3}$ and  $ \text{Br}(H_0 \to WWZ) = 7.45 \times 10^{-4}$.
 From Figs.~\ref{branchings}(a) and~\ref{branchings}(b), we find that as $f$ takes values closer to 3000 GeV, the magnitudes of the branching ratios are slightly larger. Thereby, $ \text{Br}(H_0 \to X)$ is sensitive to variations in the energy scale $ f $.
 The above effect does not happen when we vary $ \text{Br}(H_0 \to X)$ versus the $ F $ scale; this is because, in the study scenario of our choice, the condition $c_g=s_g$ removes from the interaction vertices (in most of them) the dependence on the $F$ scale.

\subsection{ Higgs boson production $ H_0 $ of the BLHM at the LHC and the FCC-hh }

We present an approximate study of the production cross-section of the $H_0$ Higgs in the BLHM, which have decay channels $\gamma \gamma$, $\gamma Z$, and $g g$. For this purpose, we employ the Breit-Wigner resonant cross-section~\cite{pdg:2023}. In this approximation, the production cross-section via gluon fusion can be calculated as follows,

\begin{eqnarray}
  \sigma(gg\to H_0\to Y) &=& \frac{\pi}{36}\frac{Br(H_0\to gg) Br(H_0\to Y)}{m_{H_0}^2},
\end{eqnarray}

\noindent where $Y= \gamma\gamma, \gamma Z,gg$. The cross-section $ \sigma(gg\to H_0\to Y)  $ is determined just at the resonance of the $ H_0 $ Higgs. Although the method of analysis proposed in this subsection approximates the production mechanism of a massive scalar particle via gluon fusion, it could provide experimental guidance for the search for new heavy particles of the TeV order.

\begin{figure}[H]
\center
\includegraphics[width=8.10cm]{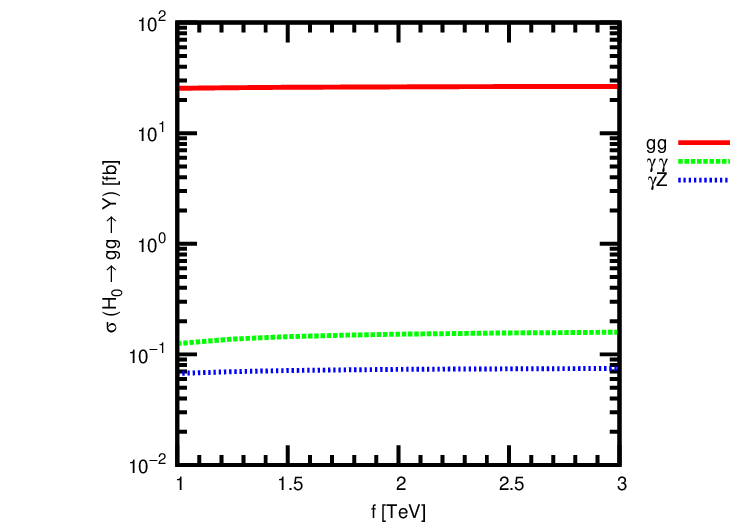}
\caption{ \label{eficazb} The production cross-section of the $H_0$ Higgs via gluon fusion as a function of the energy scale $f$ (with the fixed value of $F=4000$ GeV).}
\end{figure}

\begin{figure}[H]
\center
\subfloat[]{\includegraphics[width=8.10cm]{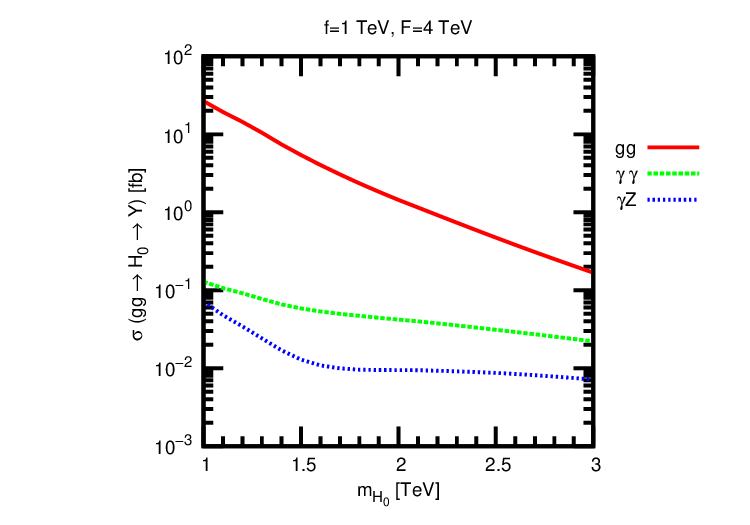}}
\subfloat[]{\includegraphics[width=8.10cm]{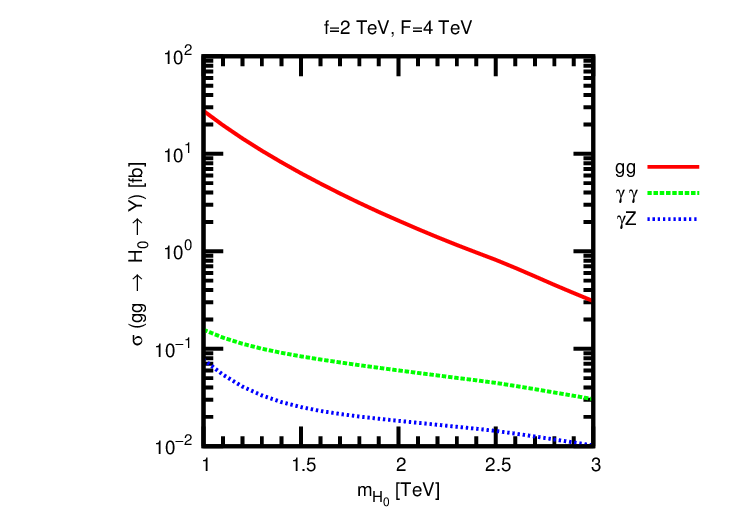}}
\caption{ \label{eficazbmH0} The production cross-section of the $H_0$ Higgs via gluon fusion as a function of  $m_{H_0}$  with a) $ f=1 $ TeV and b) $ f=2 $ TeV. }
\end{figure}

Based on previous studies where the sensitivity of the $H_0$ partial decay widths and branching ratios on the $F$ energy scale has been analyzed, the numerical estimates suggest that both $\Gamma(H_0\to X)$ and $\text{Br}(H_0\to X)$ show almost negligible sensitivity to the $F$ energy scale. Thereby, we compute the $H_0$ production cross-section only as a function of the parameter $f$ as shown in Fig.~\ref{eficazb}.
In this figure, we observe that the curve providing the largest contribution is generated by the production cross-section of $H_0$ with $ gg $ final states,
 $\sigma(gg\to H_0\to gg)= [25.45, 26.48]$ fb when $ f=[1000,3000] $ GeV.
On the other hand, the weakest contributions arise for
$\sigma(gg\to H_0\to \gamma\gamma)=[1.25, 1.59]\times 10^{-1}$ fb and  $\sigma(gg\to H_0\to \gamma Z)=[6.72, 7.46]\times 10^{-2}$ fb.
Additionally, we discuss the behavior of $ \sigma(gg\to H_0\to Y) $ vs. $m_{H_0}$ as can be seen in Fig.~\ref{eficazbmH0}. For the energy scale $ f $, we assign fixed values such as 1000 GeV and 2000 GeV. With these input values for $ f $, we generate the curves shown in Figs.~\ref{eficazbmH0}(a) and~\ref{eficazbmH0}(b).  Based on the corresponding figures and the numerical estimations, we obtain that the dominant contribution to $ \sigma(gg\to H_0\to Y) $ is reached through the $  H_0\to gg $ decay channel when $f=1000$ GeV, its production cross-section is $ \sigma(gg\to H_0\to gg)=[26.73, 1.68\times 10^{-1}] $ fb for $ m_{H_0}=[1000,3000] $ GeV.
 The subdominant contribution emerges when $f=2 000$ GeV being  $ \sigma(gg\to H_0\to gg)=[26.61, 3.08\times 10^{-1}] $ fb. The other curves generate slightly more suppressed contributions than the main contribution; 
$ \sigma(gg\to H_0\to \gamma \gamma)=[1.29 \times 10^{-1}, 2.21\times 10^{-2}] $ fb and $ \sigma(gg\to H_0\to \gamma Z)=[7.15 \times 10^{-2}, 7.14 \times 10^{-3}] $ fb while $f=1000$ GeV, and $ \sigma(gg\to H_0\to \gamma \gamma)=[1.57 \times 10^{-1}, 3.04 \times 10^{-2}] $ fb and $ \sigma(gg\to H_0\to \gamma Z)=[7.75 \times 10^{-2}, 1.02 \times 10^{-2}] $ fb when $f=2000$ GeV.

 In the experimental scenario, the heavy Higgs production mechanism such as the $H_0$ Higgs of the BLHM could be studied in the context of the LHC and its upgrades (High Luminosity (HL)-LHC, High Energy (HE)-LHC)  or at future colliders like the FCC-hh.
 After LHC Long Shutdown 2, the expected integrated luminosity of the LHC in Run 3 is approximately 450 fb$^{-1}$~\cite{cern:2022}. On the other hand,
  the HL-LHC~\cite{Abada:2019HE,Abada:2019FCC,Abada:2019FCC2} is planned to operate at a center-of-mass energy of 14 TeV with an integrated luminosity of $3\, 000$ fb$^{-1}$ while the HE-LHC~\cite{Abada:2019HE,Abada:2019FCC,Abada:2019FCC2} would provide $p p$ collisions with a center-of-mass energy of 27 TeV and an integrated luminosity of 10 000 fb$^{-1}$.  In the future experiment, the FCC-hh~\cite{Abada:2019FCC,Abada:2019FCC2,Barletta:2014vea} is designed to collect a total integrated luminosity of 30 000 fb$^{-1}$ operating at a center-of-mass energy of 100 TeV.
Considering the production cross-sections $  \sigma(gg\to H_0\to Y) $ and the expected integrated luminosities of the colliders mentioned above, we can obtain an estimate of the number of events that could be observed at the colliders for the processes of interest.
For the purpose of generating a benchmark,  considering $m_{H_0} \approx 1000$ GeV,  we provide in Tables~\ref{Ev3b}-\ref{Ev2b}  the expected events related to $H_0 \to gg, \gamma \gamma, \gamma Z$ decays when the scale of the new physics $ f $ takes specific values such as 1000, 2000, and 3000 GeV.
According to the numerical data, the one-loop decay channel of the $H_0$ Higgs corresponding to two gluons would be of great interest for the search of the hypothetical heavy particle, the $H_0$ Higgs of the BLHM.
Concerning the $H_0 \to \gamma \gamma, \gamma Z$ decays, their respective expected event magnitudes also promise a very optimistic scenario as they appear to be within the measurement range of the previously proposed future colliders.

\begin{table}[H]
\center
\caption{The number of expected events related to $H_0\to gg$ decay.
\label{Ev3b}}
\begin{tabular}{|c|c|c|c|c|c|c|}\hline\hline
\multirow{3}{*}{$f$ [TeV]} & \multicolumn{4}{c|}{\textbf{Number of expected events at the colliders:}} \\
  \cline{2-5}
& LHC & HL-LHC   & HE-LHC    & FCC-hh\\
& $\mathcal{L}=450$ fb$^{-1}$ & $\mathcal{L}=3\, 000$ fb$^{-1}$   & $\mathcal{L}=10\, 000$ fb$^{-1}$   & $\mathcal{L}=30\, 000$ fb$^{-1}$ \\
\hline
1 & 11\,453 & 76\, 355 &  254\, 516 & 763\, 548 \\
2 & 11\,809 & 78\, 731 &  262\, 438 & 787\, 315 \\
3 & 11\,914 & 79\, 431 &  264\, 769 & 794\, 308  \\
\hline
\hline
\end{tabular}
\end{table}

\begin{table}[H]
\center
\caption{The number of expected events related to $H_0\to \gamma\gamma$ decay.
\label{Ev1b}}
\begin{tabular}{|c|c|c|c|c|c|c|}\hline\hline
\multirow{3}{*}{$f$ [TeV]} & \multicolumn{4}{c|}{\textbf{Number of expected events at the colliders:}} \\
  \cline{2-5}
& LHC & HL-LHC   & HE-LHC    & FCC-hh\\
& $\mathcal{L}=450$ fb$^{-1}$ & $\mathcal{L}=3\, 000$ fb$^{-1}$   & $\mathcal{L}=10\, 000$ fb$^{-1}$   & $\mathcal{L}=30\, 000$ fb$^{-1}$ \\
\hline
1 & 56 & 376 &  1\, 252 & 3\, 756 \\
2 & 68 & 457 &  1\, 523 & 4\, 568 \\
3 & 71 & 476 &  1\, 587 & 4\, 760  \\
\hline
\hline
\end{tabular}
\end{table}

\begin{table}[H]
\center
\caption{The number of expected events related to $H_0\to \gamma Z$ decay.
\label{Ev2b}}
\begin{tabular}{|c|c|c|c|c|c|c|}\hline\hline
\multirow{3}{*}{$f$ [TeV]} & \multicolumn{4}{c|}{\textbf{Number of expected events at the colliders:}} \\
  \cline{2-5}
&LHC & HL-LHC   & HE-LHC    & FCC-hh\\
& $\mathcal{L}=450$ fb$^{-1}$ & $\mathcal{L}=3\, 000$ fb$^{-1}$   & $\mathcal{L}=10 \, 000$ fb$^{-1}$   & $\mathcal{L}=30\, 000$ fb$^{-1}$ \\
\hline
1 &30  & 202 &  672 & 2\, 017 \\
2 &33  & 220 &  732 & 2\, 195 \\
3 &34  & 224 &  746 & 2\, 238  \\
\hline
\hline
\end{tabular}
\end{table}

\section{Conclusions} \label{conclusions}

In this work, we perform a phenomenological study of the production of the heavy Higgs boson $H_0$ via gluon fusion in the context of the BLHM.
Specifically, we analyze the one-loop decays of the $H_0$ Higgs, which refer to $ H_0 \to gg, \gamma \gamma, \gamma Z $ processes. For these decays, the effects induced by the new particles of the BLHM and the particles of the SM are considered. As the BLHM has two independent energy scales, $f$ and $F$, these represent the scales of the new physics of the model. In this way, we have generated phenomenological results for the branching ratios and production cross-sections of the $ H_0 $ Higgs in analysis regions corresponding to  $f=[1000,3000]$ GeV and $F=[3000,6000]$ GeV. For the considered intervals of the scales $f$ and $F$, we analyzed the dependence of $\text{Br}(H_0 \to X)$  on the aforementioned scales and found that the branching ratio shows sensitivity to variations in the $ f $ scale, this effect is not observed with the $F$ scale. In the two study scenarios, the dominant branching ratios at the tree and one-loop level correspond to processes $H_0 \to \bar{t}t$ and $H_0 \to gg$ whose numerical predictions are of $10^{-1}$ and $ 10^{-2} $, respectively.
A rough estimate of the production cross-section of $H_0$ was also implemented via gluon fusion. For this case, the numerical estimates of $\sigma(gg\to H_0\to Y)$
tell us that the $H_0 \to gg$ process offers a very promising scenario for the search of the heavy particle $H_0$ in future experiments such as LHC, HL-LHC, HE-LHC, and FCC-hh. In this approach, we have that for $m_{H_0} \approx 1000$ GeV and
 $f=1$ TeV we could estimate around $11\, 453$ events at the LHC, $76\, 355$ events at the HL-LHC, $254\, 516$ events
at the HE-LHC and $763\, 548$ events for the FCC-hh, which is a very optimistic scenario for the study of the scalar $H_0$ predicted by the BLHM.

\vspace{4.0cm}

\begin{center}
{\bf Acknowledgements}
\end{center}

E. C. A. appreciates the post-doctoral stay. A. G. R. and D. E. G.  thank SNII (M\'exico).

\newpage

\end{document}